# Re-entrant melting and freezing in a model system of charged colloids

C. Patrick Royall[1], Mirjam E. Leunissen[2], Antti-Pekka Hynninen[2], Marjolein Dijkstra[2] and Alfons van Blaaderen[2]

[1]Present address: Institute of Industrial Science, University of Tokyo, Komaba 4-6-1, Meguro-ku,Tokyo 153-8505, Japan

[2]Soft Condensed Matter, Debye Institute, Utrecht University, Princetonplein 5, 3584 CC Utrecht, The Netherlands

We studied the phase behavior of charged and sterically stabilized colloids using confocal microscopy in a less polar solvent (dielectric constant 5.4). Upon increasing the colloid volume fraction we found a transition from a fluid to a body centered cubic crystal at 0.0415±0.0005, followed by re-entrant melting at 0.1165±0.0015. A second crystal of different symmetry, random hexagonal close-packed, was formed at a volume fraction around 0.5, similar to that of hard spheres. We attribute the intriguing phase behavior to particle interactions that depend strongly on volume fraction, mainly due to changes in the colloid charge. In this low polarity system the colloids acquire charge through ion adsorption. The low ionic strength leads to fewer ions per colloid at elevated volume fractions and consequently a density-dependent colloid charge.

I. Introduction

Freezing and melting are common everyday physical phenomena. More unusual is re-entrant melting, which usually stems from a subtle interplay between enthalpy and entropy, and may be found in systems as diverse as discotic liquid crystals ,[1] diblock copolymer solutions [2] and Helium-3 .[3] We present results on a model system of charge-stabilized colloids, which exhibits an intriguing re-entrant colloidal fluid phase at a higher volume fraction than a colloidal crystal, and a second crystal phase with different symmetry at a higher volume fraction still. The phase behavior of colloids dispersed in a solvent is thermodynamically equivalent to that of atoms and small molecules,[4,5] however colloids can be studied with optical microscopy due to their relatively large size. We analyzed the structure in 3D real space, at the single particle level using confocal laser scanning microscopy (CLSM).[6]

Under many circumstances, the biologically and industrially relevant system of charged colloids is described by a screened Coulomb ('Yukawa') interaction.[7,8,9] In this work we will neglect the Van der Waals attractions, since they are reduced to a fraction of the thermal energy by matching the refractive index of the colloids and the solvent.[5] A linearization of the mean-field Poisson-Boltzmann theory used by Derjaguin, Landau, Verwey and Overbeek,[7] combined with a hard-core term due to the physical size of the colloids, describes the interaction between two colloids with constant surface potential as

$$\beta u(r) = \begin{cases} \infty & r \leq \sigma \\ \dfrac{Z^2}{(1+\kappa\sigma/2)^2} \dfrac{l_B}{\sigma} \dfrac{\exp(-\kappa(r-\sigma))}{r/\sigma} & r > \sigma \end{cases} \quad (1)$$

where $\beta = 1/k_B T$, $Z$ is the number of charges on a colloid, $\sigma$ is the colloid diameter, $r$ is the center-to-center separation, $k_B$ is Boltzmann's constant and $T$ is the absolute temperature. The Bjerrum length of the solvent is given by $l_B = \beta e^2 / \varepsilon_S$ where $e$ is the elementary charge and $\varepsilon_S$ is the permittivity. The inverse Debye screening length is denoted by $\kappa = \sqrt{4\pi l_B \rho_i}$ where $\rho_i$ is the total number density of monovalent ions. These ions have several sources, 'background' ions, due to solvent self-dissociation and impurities, and counter-ions which balance the charge on the colloids.

Often the parameters used in equation (1) are taken to be constant with respect to volume fraction. In earlier work we found this to be a good approximation.[10] However, as we shall discuss in this paper, the colloid charge and Debye length can be a function of volume fraction for a variety of reasons. The linear Poisson-Boltzmann theory assumed in equation (1) is only valid for small electrostatic potentials (colloid charges). However, at higher potentials, the pair interaction can still be described as Yukawa-like, but with a smaller, renormalized charge, except for very small separations.[8] Charge renormalization leads to volume fraction dependent interactions.[11] Renormalization has successfully been used to compare experimental phase behavior to the Yukawa potential,[12] but did not explain a re-entrant fluid formed by increasing the colloid charge at constant volume fraction.[13] In addition, when the counter-ions form a significant fraction of the ionic strength, the colloid-colloid interactions depend strongly on the volume fraction: decreasing the volume fraction reduces the ionic strength, increases the Debye length and can enhance the structure at low densities.[14]

Moreover, when the range of the interaction exceeds the mean interparticle separation, the assumption of pairwise addivity becomes questionable [15,16]. Effective attractions due to non-pairwise additivity [17] may [18] help explain a range of recent experimental observations such as superheated crystals,[19] and 'liquid-gas' phase separation,[20] which are not expected for purely repulsive interactions. Attractions can also be caused by correlations between small ions,[21] Volume fraction dependent attractions may also explain re-entrant phase behavior, so in addition to simulations using equation (1), we compared our results with the primitive model. Non-pairwise addivity and correlations are included in the primitive model, where all charged species, including the small ions, interact via a Coulomb potential with a hard core:

$$\beta u(r) = \begin{cases} \infty & r < \frac{1}{2}(\sigma_i + \sigma_j) \\ \frac{q_i q_j l_B}{r} & r \geq \frac{1}{2}(\sigma_i + \sigma_j) \end{cases} \quad (2)$$

where $i$ and $j$ are the interacting species, and may be colloids ($q=Z$) or monovalent co- or counter-ions ($q=1$) and $\sigma_{i/j}$ is the diameter.[22,23]

Here we studied the phase behavior of a suspension of charged and sterically stabilized poly-methyl methacrylate (PMMA) colloids in a in an (almost) index and density matching solvent mixture with a dielectric constant of 5.4 .[24] Upon increasing the volume fraction, we found a colloidal fluid, followed by a body centered cubic (BCC) crystal, which melted to a re-entrant fluid and then a second freezing transition to a crystal of different symmetry, random hexagonal close packed (RCHP) (Fig. 1). By contrast, the phase diagram for Yukawa systems (Eq.(1)) exhibits only one fluid-crystal transition, as a function of colloid concentration.[34] The re-entrant melting observed here may be reconciled with the Yukawa interaction only by allowing the colloid charge and Debye screening length to vary as a function of volume fraction. In other words, the Yukawa interaction is density-dependent. By comparing the results of experiments with Monte Carlo (MC) simulations, we can obtain an estimate of the colloid charge and Debye length as a function of volume fraction. We also made independent measurements of the colloid charge in dilute dispersions with electrophoresis. The work reported here contrasts with previous measurements on a similar system,[10] in which we found a fixed pair potential as a function of colloid volume fraction. We highlight the differences between these experiments below. This paper is organized as follows: first we describe our experimental and simulation methods, followed by results. Our discussion identifies possible causes for the unusual phase behavior observed.

II. Experimental

We used sterically stabilized, poly-methyl methacrylate colloids of 2.16 $\mu$m diameter in a solvent mixture of cis-decalin and cyclohexyl bromide (CHB). By closely matching both the refractive index and mass density of the colloids,[24] the system is optimized for optical techniques. The colloids were labeled with rhodamine isothiocyanate [25] but we saw similar behavior using colloids labeled with 7-nitrobenzo-2-oxa-1,3-diazol, and conclude that the dye has little influence. We purified the CHB solvent by washing and distilling as described previously.[10] The dielectric constant of the solvent mixture was 5.37, determined with dielectric spectroscopy.[26] The dispersions were confined to glass capillaries with inner dimensions of 0.1 x 1.0 mm (VitroCom) and sealed at each end with Norland Optical adhesive no 68 (Norland Optical Products Inc). We checked that melt-sealing the glass gave the same behavior. In a typical experiment the dispersion had an initial volume fraction of 0.04 and was centrifuged at 1000 rpm (Hettich Zentrifugeren Rontina 46s) for 18 hours to create a concentration gradient, and then lay horizontally for 24 hours to allow the small ions to equilibrate. After sedimentation we found that around 90% of the capillary was free from colloids, and functioned as an ion reservoir.

To reveal the phase behavior we scanned the gradient in colloid concentration resulting from the centrifugation using a Leica SP2 CLSM with a 63x NA 1.4 oil immersion objective, in fluorescence mode with 543 nm excitation. The images were taken at least 20 $\mu$m from the capillary wall. Recently, it was shown that charged colloids in a gravitational or centrifugal field can set up a macroscopic electric field,[27] but the impact of this electric field on the pair interactions is negligible for our range of parameters.[28] We stress that, although the system is in principle in a metastable state due to the concentration gradient, the diffusion of 2.16 $\mu$m diameter colloids is sufficiently slow that the gradient, which typically spanned a few millimetres, was still present after months. Certainly, no local change in colloid volume fraction and structure was found during data acquisition (up to 3 hours). Although the colloid concentration gradient may be regarded as fixed, the small ion concentration is expected to relax after a few hours, on a centimeter length-scale. We present results from concentration gradient samples in equilibrium with an ion reservoir, but also prepared a number of separate samples at differing volume fractions, and saw similar behavior, giving strong evidence that our observations are not due to the metastable density profile induced in the colloids by centrifugation.

We analysed the structure of the colloidal suspension to obtain a value for the colloid charge. Other, more direct methods may also be used in principle, such as titration, electrophoresis and electroacoustic characterization. However, these techniques do not necessarily determine the effective charge that governs the experimental phase behaviour. For instance, titration measures the bare charge, which is not relevant here. The electrokinetic techniques have another drawback: their theoretical interpretation at higher volume fractions is not sufficiently developed at this time. However, we have performed electrophoresis measurements on dilute ($\eta$=0.0015) samples. We made the measurements with a Coulter Delsa 440SX and determined the stationary layers from Komagata linearisation. We used the Hückel equation to relate the mobility to the zeta potential and used the empirical relationship proposed by Loeb *et al*. to relate the zeta-potential to the colloid charge.[29] We assumed that the zeta-potential is the actual surface potential, which is appropriate in the case of small surface charge and long Debye length, as is the case here. We also made conductivity measurements using a variety of solvent batches and colloid concentrations, using a Scientifica 627 conductivity meter, in order to study the origin of the re-entrant melting behaviour.

III. Analysis: Determining the colloid charge and Debye screening length

Central to this work is the extraction of the colloidal interactions from confocal microscopy data. We consider two strategies, for colloidal fluids and crystals respectively. In the case of fluids, a measurement of the radial distribution function $g(r)$ can be used to determine the *pair* interaction. We found that the assumption of pairwise additivity is indeed a sufficient description of this system. In principle, the interaction potential is uniquely determined by the $g(r)$ for pairwise additive systems,[30] and may be extracted by inversion techniques [31]. However, such methods require a radial distribution function of higher precision than that which we can measure here in 3D (the details of our method to determine $g(r)$ are given in Ref. [10]). We therefore assumed that the colloidal interactions have a Yukawa form (Eq.(1)). We carried out MC simulations in the canonical ensemble [32] with various combinations of the colloid charge $Z$, and the Debye screening length $\kappa^{-1}$, and compared the resulting $g(r)$'s with the experimental data. We selected those which gave the best agreement and assumed the input parameters to the simulation were the same as those in the experiment. However, due to the error in the measured radial distribution functions, there was a range of input parameters to the simulation which gave good agreement, thus there is an uncertainty in the values we obtained for the colloid charge and Debye length, which depends on the state point. The uncertainty is estimated as 20% in both $Z$ and $\kappa^{-1}$ for the dilute fluid phase, and 20% in $Z$ and 40% in $\kappa^{-1}$ for the re-entrant fluid phase. In every experiment, the values of $Z$ and $\kappa^{-1}$ were corresponded to a fluid according to equation (1).

We extended our analysis to address some of the assumptions of equation (1). In particular, the restrictions of linear Poisson-Boltzmann theory and pairwise additivity between the colloids are lifted in the primitive model (Eq.(2)). The primitive model also includes effects of ion correlations, although these are extremely small in this system. In the primitive model simulations, we again varied the colloid charge and Debye length and compared the results both to the Yukawa interaction (Eq.(1)) and experiment. Here the primitive model is implemented on a lattice, where the lattice spacing is chosen so each colloid diameter is divided into 19 lattice sites.[23] We found that different sized lattices gave indistinguishable $g(r)$'s for our parameters, and concluded that the lattice discretization had no effect on our results.

In the colloidal crystal, the particles are confined to a potential well formed by the interactions with their neighbours, $w(d)$, where $d$ is the displacement from the lattice site. For displacements less than 10% of the interparticle separation, the well is approximately harmonic, so $\beta w(d) \approx \alpha d^2/2$ with $k$ the 'effective spring constant', as shown in Fig. 2.[33] To calculate $\alpha$, we sample typically 500 configurations of around 400 particles. The mean position of each particle is determined, and the excursions are measured, from which we calculate the likelihood of an excursion of given distance. The logarithm of the likelihood then gives the potential well, which we fit to a quadratic form for less than 10% of the interparticle separation (fig. 2). In exactly the same way, we can also calculate $\alpha$ from simulation data. Thus, given a trial pair interaction, we can compare the results of simulated and experimental 'effective spring constants', in a conceptually similar way to the $g(r)$ method above. However, since we compare only a scalar, fitting a 2-parameter potential such as equation (1) is not possible. We can simply identify whether a given input potential gives good agreement with the experimental data.

IV. Results

Previously, we found a fluid to face centered cubic crystal phase transition at a colloid volume fraction of $\eta \approx 0.1$.[10] Both the phase transition and the fluid structure were consistent with a density independent Yukawa potential (Eq. (1)) [34] with a colloid charge around 600 and Debye screening length of 400 nm. Those results contrast strongly with the behavior shown in Fig. 1 which forms the subject of this article. For $\eta \leq 0.041$, a colloidal fluid is seen (Fig. 1(*a,b*)), whose long-ranged repulsions are suggested by the considerable separation between the particles in Fig. 1(*a*) (see also Fig. 3(*a*), bottom line). Upon compression this fluid forms a BCC crystal at $\eta = 0.0415 \pm 0.0005$ (Fig. 1(*c*)). What is unusual is that the crystal melts, at $\eta = 0.1165 \pm 0.0015$ (Fig. 1(*d, e*)) into a re-entrant fluid (Fig. 1(*e*)), with completely isotropic diffusion of the particles (Fig. 1(*f*)). There is a second crystallization, to an RHCP structure, at $\eta \approx 0.50$, (Fig. 1(*g*)). This second crystallization cannot be distinguished from that reported for hard spheres [4], although the colloids still carry some charge.[34]

Figure 3 shows experimental radial distribution functions and the results from the Yukawa potential and primitive model simulations. In the low volume fraction fluid (Fig. 3(*a*)), we fitted the $g(r)$s with a modest decrease in charge from $Z=412$ to $Z=358$ from $\eta = 0.0037$ to 0.041. The charge reduction in the re-entrant fluid is rather greater, to 185 by $\eta = 0.118$ and 117 by $\eta = 0.187$ (Fig. 3(*b*)). In the low volume fraction fluid ($\eta \leq 0.041$), the Debye length fell from 1.4±0.3 $\mu$m to 1.1±0.2 $\mu$m, while in the re-entrant fluid, g(r) fitting suggests $\kappa^{-1}$=1.0±0.4$\mu$m. These values are consistent with ions in a colloidal suspension in 'Donnan' equilibrium with an ion reservoir with $\kappa^{-1}$=1.5 $\mu$m at the colloid charge and volume fraction measured. The results from the primitive model have the same trend of a charge which falls with volume fraction, but the values are slightly lower, Z=320 and 90 for $\eta$=0.041 and 0.187 respectively, in line with non-linear Poisson-Boltzmann results.[9] Since the primitive model, which includes non-pairwise additivity and is not restricted to the linear Poisson-Boltzmann regime, gives similar values to the Yukawa interaction, we conclude that neither of these effects matter greatly for our parameters. Within the scope of this work, the pairwise Yukawa interaction (equation (1)) thus provides a satisfactory description. Although the reduction in charge is consistent with a re-entrant fluid, on the basis of the phase behavior alone we cannot exclude the possibility of a re-entrant fluid phase in the presence of a constant colloid charge: if the Debye length were to fall sufficiently, then a Yukawa model could predict re-entrant melting as well.[34] However, in this case, a threefold fall in the Debye length is needed in the range $\eta = 0.0415 \pm 0.0005$ to $\eta = 0.1165 \pm 0.0015$, corresponding to an unphysical order of magnitude increase in ionic strength. We have also measured the conductivity as a function of colloid volume fraction, and find no large increase in conductivity, further supporting our analysis of the confocal microscopy data.

The decrease in colloid charge with volume fraction is shown in Fig. 4. Unfilled circles are obtained from g(r) fits with the Yukawa potential, while triangles are primitive model results. In the BCC phase, we were able to estimate the Debye length by interpolating between the values at the highest density in the dilute fluid phase ($\eta$=0.041, $\kappa^{-1}$=1.1) and the lowest density in the re-entrant fluid phase ($\eta$=0.118, 1.0 $\mu$m). In order to determine the colloid charge, we further assumed sufficiently strong interactions for crystallization according to equation (1).[34] This provided a lower bound to the colloid charge. As mentioned above, in the BCC phase, we determined an 'effective spring constant' $k$. For these parameters, the potential well which confines each particle is almost spherically symmetric. Consequently, similar values of $k_x$ (triangle right), and $k_y$ (triangle up) (Fig. 4, inset) were obtained. The directions $x$ and $y$ are defined by the orientation in Fig. 1(*d*). We also determined $k$ from Yukawa potential MC simulations, using the values we estimated above for the charge in the BCC crystal phase (Fig. 4, filled squares). The simulated $k$ values are shown as filled squares in the inset of figure 4. The agreement with the experimental data is good. For comparison, we also assumed a constant charge of $Z$=358 (Fig. 4 inset, unfilled squares). In this constant charge case, $k$ increases as a function of volume fraction significantly faster than the experimental measurements. Fig. 4 (inset) shows that in the colloidal crystal too, the charge appears to decrease. We also show the results of electrophoresis measurements, $Z$=560±50, (Fig. 4, filled circle). This provides an independent check of the colloid charge at low volume fractions which is consistent with our analysis of the confocal microscopy data.

Yukawa systems with arbitrary Debye lengths, volume fractions and colloid charges may be plotted on a single set of axes in the $(\lambda, \tilde{T})$ [35] representation, where $\lambda$ is a scaled Debye length and $\tilde{T}$ is an effective temperature. Furthermore, in this representation, the phase boundaries are almost straight lines.[36] Although the $(\lambda, \tilde{T})$ representation only applies to *point* Yukawa particles, it turns out that, for the parameters considered here, the hard core has negligible effect on the phase behavior for $\eta \leq 0.187$.[34] We therefore ignore the hard core, and map the parameters of our statepoints obtained from fitting the experimental data with a Yukawa interaction to $(\lambda, \tilde{T})$ as follows:

$$\lambda = \kappa\sigma(6\eta/\pi)^{-1/3}$$

(3)

$$\widetilde{T} = \left[\frac{2}{3}\lambda^2 \beta u_M(\lambda)\right]^{-1}$$

(4)

where $u_M$ is the Madelung energy per particle in an ideal FCC crystal.[34] Fig. 5 shows the statepoints of the re-entrant system that is the subject of this article (unfilled symbols) and the previous statepoints which did not show any re-entrant melting (filled symbols).[10] The non-re-entrant system takes only a slightly curved path as the colloid volume fraction is increased, as indicated by the arrows. Since the melting line is approximately straight in fig. 5, in order to cross the line twice (re-entrant melting), the path in the ($\lambda$, $\widetilde{T}$) representation must exhibit a return. Our experimental statepoints show good agreement with the phase diagram of point Yukawa particles.

V. Conductivity measurements

In the above, we have argued that the colloid interactions depend strongly on volume fraction. In particular, both the primitive model and Yukawa potential show that the colloid charge appears to decrease with volume fraction. This fall is too large to be explained by assuming a constant surface potential, rather than constant charge.[7]

The exact nature of the charging process in the solvent mixture is not yet fully understood, so providing a complete explanation of the fall in colloid charge is difficult. Here we present some observations that may guide towards an understanding of the chemistry underlying the unusual phase behaviour observed. We begin by summarising our observations. We have studied four systems of PMMA in the CHB-decalin solvent mixture. As mentioned above, various batches of fluorescently labeled PMMA were used. No significant differences were seen between the batches. Table (I) presents the results of the phase behaviour and conductivity measurements we carried out. In experiment (1),[10] we found a constant colloid charge as a function of colloid volume fraction. Experiment (2), the dispersion displaying re-entrant melting, forms the subject of this paper, and here we used a different batch of CHB solvent ('B'). Further experiments were carried out with a third solvent batch, 'C', which was washed only (table I). In this case, 'normal' phase behaviour was observed, i.e. there was no re-entrant melting (experiment 3). When more ions were introduced to the solvent 'C' by addition of tetrabutyl ammonium bromide salt (TBAB), we again saw re-entrant melting (experiment 4).

Recent work[37] shows that HBr is present due to decomposition of the CHB solvent and suggests that the colloids acquire their charge by *adsorption* of $H^+$ ions. In the case of experiment (1), the relatively high conductivity suggests that an excess of $H^+$ was present at colloid volume fractions up to 0.2 at least, and that the colloids could acquire a charge of several hundred elementary charges without significantly altering the overall ionic concentration. Here we expect that both colloid charge and Debye length stay fairly constant as a function of colloid volume fraction. The phase behavior would then be consistent with a Yukawa interaction with a single set of parameters, as we indeed found.

For experiments (2-4) the conductivity was much lower and therefore the ion concentration may be strongly affected by adsorption of ions by the colloids and even by adsorption on the wall of the capillary. We used the conductivity measurements to calculate the ionic strength in the solvent mixture before and after adding colloids, using Walden's rule. This assumes that the product of the limiting molar conductance, $\Lambda_0$, and the viscosity is equal in two solvents. The limiting molar conductance in the solvent of interest is then the product of the ratio of the two viscosities and $\Lambda_0$ of the reference solvent (which is known). We used literature values for the limiting molar conductance in ethanol of 53.6 cm$^2$SMol$^{-1}$ for H$^+$ and 35.3 cm$^2$SMol$^{-1}$ for Br$^-$ and used Walden's rule to determine values of 18.5 cm$^2$SMol$^{-1}$for Br$^-$, 28.1 cm$^2$SMol$^{-1}$ for H$^+$ and 4.4 cm$^2$SMol$^{-1}$ for TBAB in CHB-decalin. We then assumed that the conductivity of the solvent was equal to the product of the ion concentrations and their limiting molar conductances. The contribution of the large, weakly charged colloids to the conductivity is very small and therefore neglected. In this way, taking measured values of the conductivity, we calculated the ionic strengths of the solvents (table I).

In the case of the batch 'B' CHB (experiment (2), table I) we found that, upon adding colloids, the conductivity was reduced by a factor of approximately four (table I). We regard this as evidence in favor of ion adsorption by the colloids. In fact, the addition of a colloid volume fraction $\eta$=0.03, with each particle adsorbing several hundred protons would account for almost all the H$^+$ ions, and result in a fall in conductivity comparable to that which we measured. At higher colloid concentrations we expect there are insufficient protons, so the charge per colloid would then be reduced, leading to the re-entrant melting behavior observed. The same reasoning can explain experiment (4) in which we were able to induce re-entrant melting by adding TBAB salt to the solvent. Although larger quantities of TBAB are known to induce inversion of the colloid charge,[38] such that the particles carry a negative charge, this does not occur until a TBAB concentration at least a hundred times higher than the 290 nMol added here. This concentration of TBAB gave a conductivity of 190 pS/cm, similar to that of the batch 'B' solvent which exhibited re-entrant melting. As in the case of experiment (2), the adsorption of H$^+$ and, here, TBA$^+$ and most likely some Br$^-$, reduces the calculated ionic strength and hence the conductivity, to an extent similar to that which we measured. However, the magnitude of the measured fall in conductivity is less in the case of experiment (4) than experiment (2), which may reflect the adsorption of the less mobile $TBA^+$ ions, instead of the faster protons (experiment (2)). This reasoning does not explain experiment (3). In this case, the conductivity measurements suggest that the maximum number of ions available for each colloid is too low for crystallization to occur at all, according to equation (1). However, assuming the colloids did indeed absorb the available $H^+$ and (less) $Br^-$ ions, we expect a 20-fold decrease in ionic strength, and a Debye length of some 4-5$\mu$m, around 4 times longer than experiment (2) studied in this paper. For this extremely long Debye length the screening of the colloid charge is very limited indeed and may lead to deviations from the behavior predicted by equation (1), possibly resulting in crystallization at lower volume fractions. While a full analysis goes beyond the scope of this work, preliminary comparisons between the primitive model and Yukawa potential suggest that the Yukawa potential does indeed underestimate the structure in this parameter range, as noted in non-linear Poisson-Boltzmann simulations.[9]

VI. Conclusions

We have presented results on a model system of charged colloids, which exhibits the unusual phase behavior of re-entrant melting and freezing. Strong evidence for the fact that non-pairwise additivity is *not* important here is given by the agreement of the simulations using the Yukawa potential and the primitive model. Instead, by analyzing our experimentally determined radial distribution functions and the particle excursions in the BCC crystal, we find that the colloid charge falls as a function of volume fraction, which can explain the observed phase behavior. At colloid volume fractions above a few percent, we believe there are insufficient positive ions per colloid, leading to a fall in colloid charge with volume fraction, and the re-entrant phase behaviour we observed. The second freezing transition at $\eta \approx 0.5$ is attributed predominantly to hard-core interactions. The high volume fraction crystal has a different symmetry (RHCP) to the BCC crystal formed at lower colloid volume fractions. Finally, we note that it is dangerous to consider only freezing at higher volume fractions and assume from this that a colloidal system has hard-sphere like behavior. It is clear that, as shown for the present system, the interactions at lower volume fractions should be characterized as well, and may depend strongly on volume fraction.

## Acknowledgements

The primitive model code is based on that of Athanassios Panagiotopoulos. It is a pleasure to thank Paul Bartlett, Maarten Biesheuvel, Jan Groenewold, Hans-Hennig von Grünberg, J. Hans Lyklema, Willem Kegel, Vladimir Lobaskin, Hartmut Löwen, Thomas Palberg and René van Roij for stimulating discussions. Willem Kegel and René van Roij are also thanked for a critical reading of the manuscript. Didi Derks is thanked for particle synthesis and solvent purification. This work is part of the Stichting voor Fundamenteel Onderzoek der Materie (FOM), which is supported by Nederlandse Organisatie voor Wetenschappelijk Onderzoek (NWO).

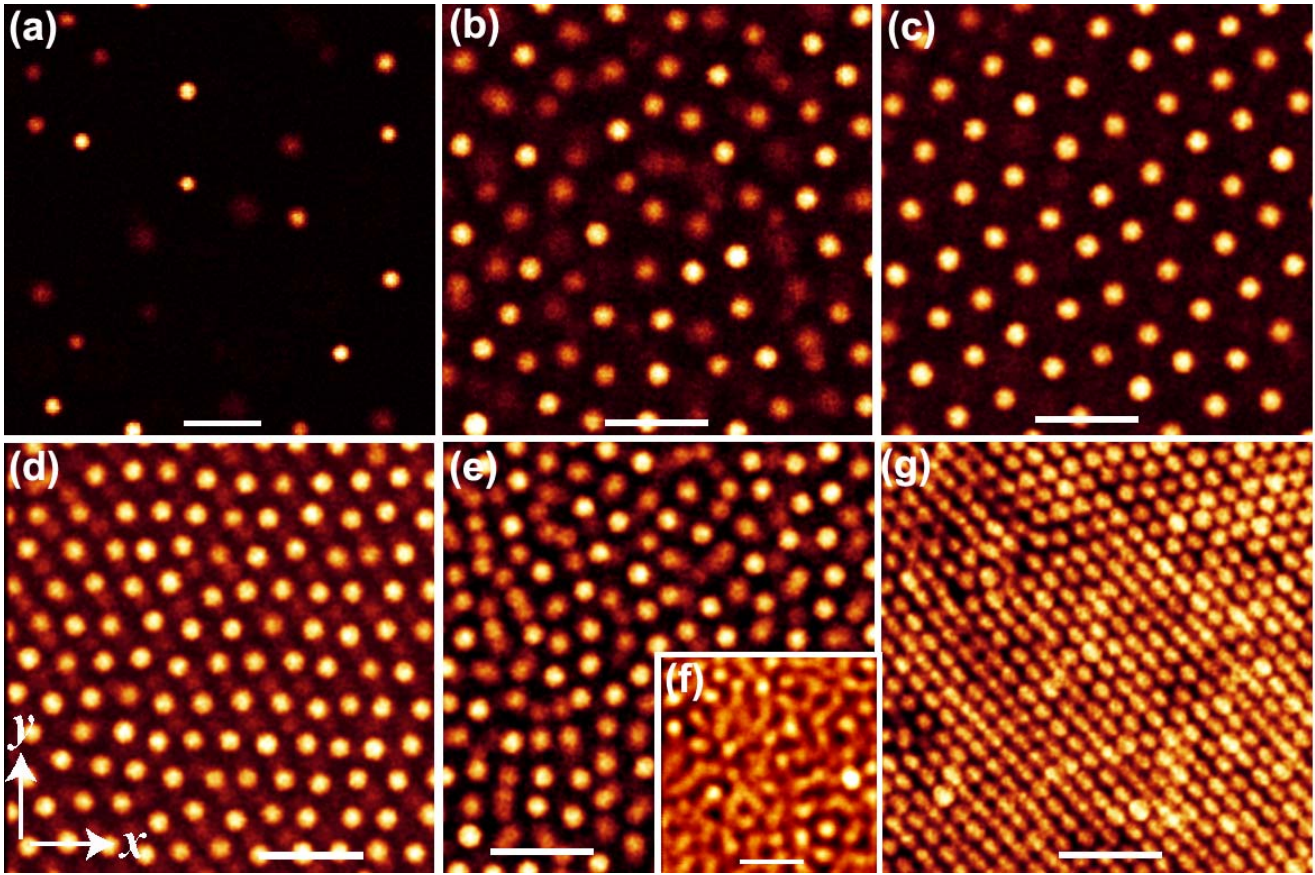

**FIG. 1: Confocal microscopy images of phase behavior as a function of volume fraction, $\eta$. Initially, a fluid is seen, (*a*, $\eta$=0.0037, *b*, $\eta$=0.041), forming a BCC crystal (*c*, $\eta$=0.042, *d*, $\eta$=0.115 ). What is unusual is the re-entrant fluid (*e*, *f*, $\eta$=0.118). Finally, a RHCP crystal is formed (*g*, $\eta\approx$0. 5). A time-lapse over four minutes, indicating homogeneous diffusion is shown in (*f*). All other images are single *xy* scans of 500 ms duration. Bars=10μm.**

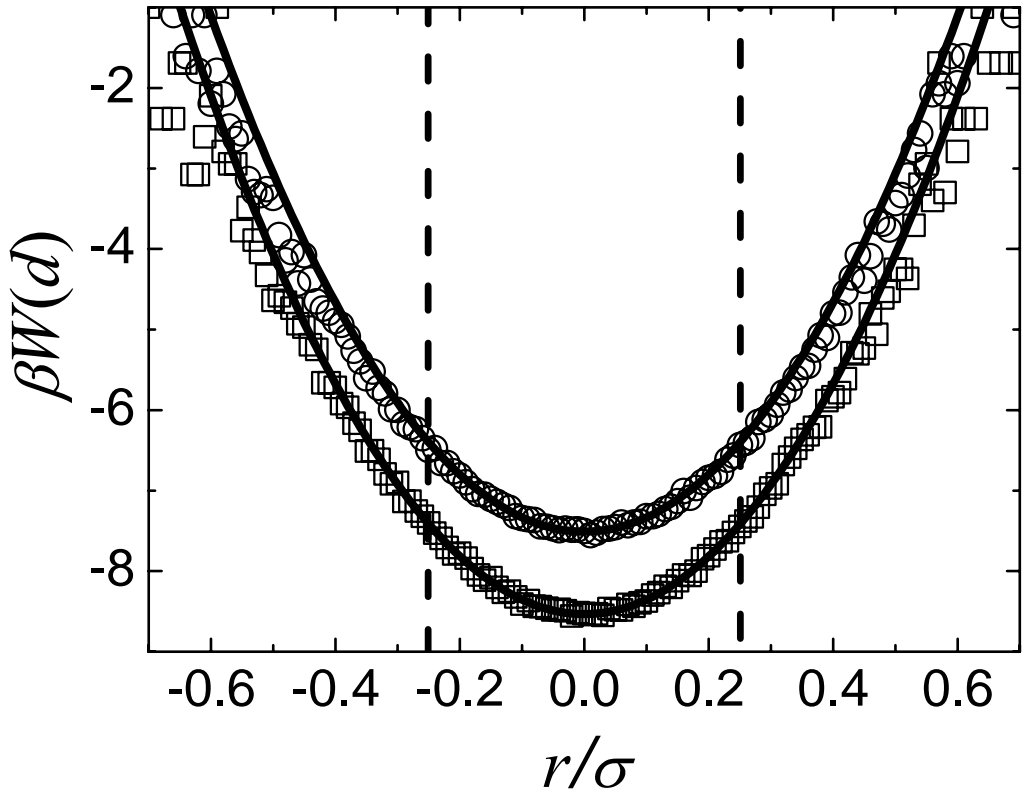

**FIG. 2. Particle Displacement from mean lattice position, for $\eta$=0.103. Squares and circles are experimental *x* and *y* data respectively. *x* and *y* are defined in figure 1(*d*). Solid lines are quadratic fits that are taken between the dotted lines to yield the 'effective spring constant', *k*. Data offset for clarity.**

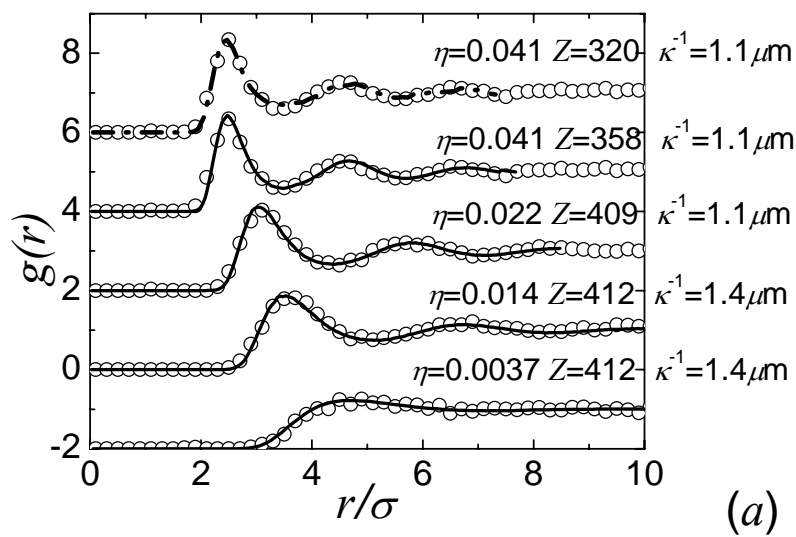

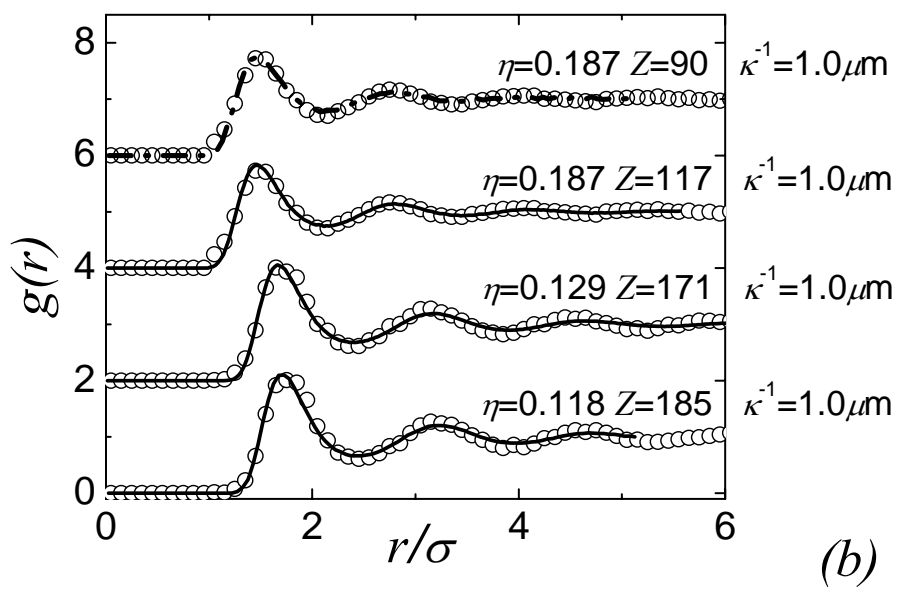

**FIG. 3. The radial distribution function at various colloid volume fractions in the low volume fraction fluid (*a*) and re-entrant fluid (*b*). Circles are experimental data. MC simulation data for the Yukawa potential (solid lines) and primitive model (top, dashed) are also plotted. Data offset for clarity.**

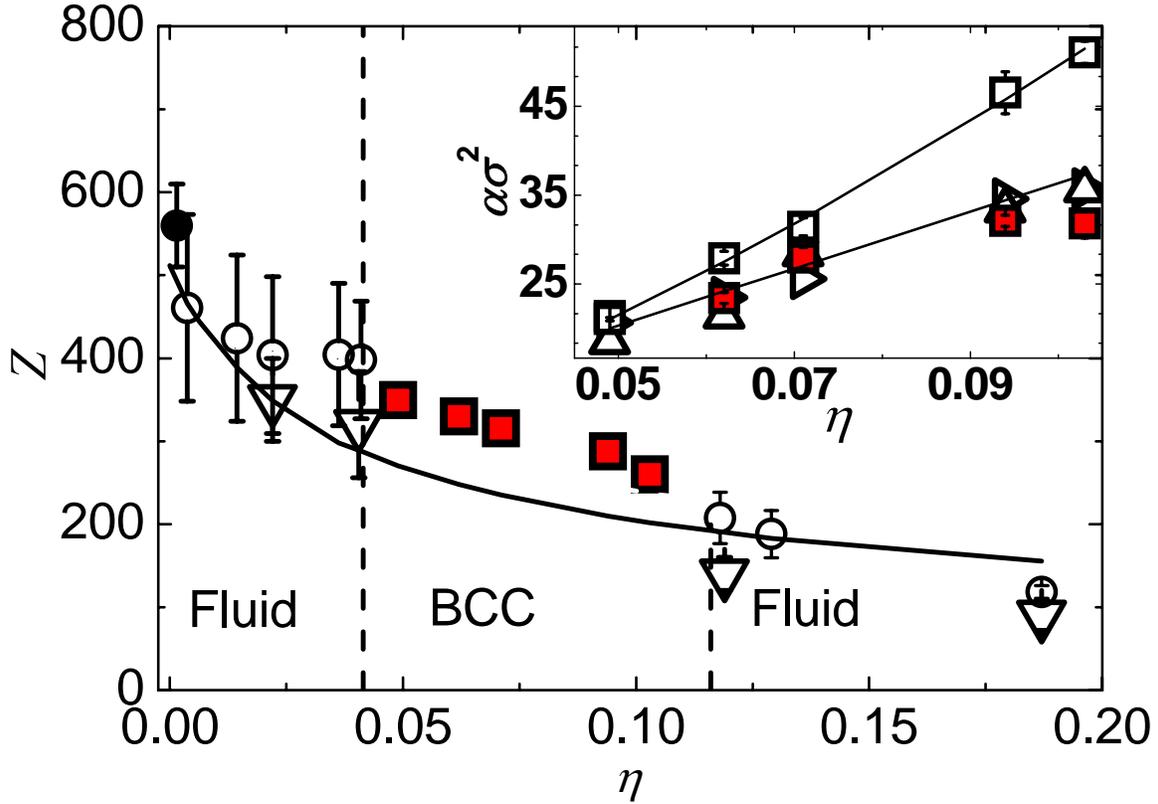

**FIG. 4 (color online). Colloid charge $Z$ plotted as a function of volume fraction, $\eta$, as found from electrophoresis (filled circle), $g(r)$ fitting with the Yukawa potential (unfilled circles) and primitive model (triangles). The filled squares denote the charge estimated from the Yukawa phase diagram [34] (filled squares). Solid line is a guide to the eye. Dashed lines are approximate phase boundaries. Inset: Effective spring constant $k$. Triangles are experimental data. Filled squares correspond to simulation using estimated charge in main figure, unfilled squares to constant charge.**

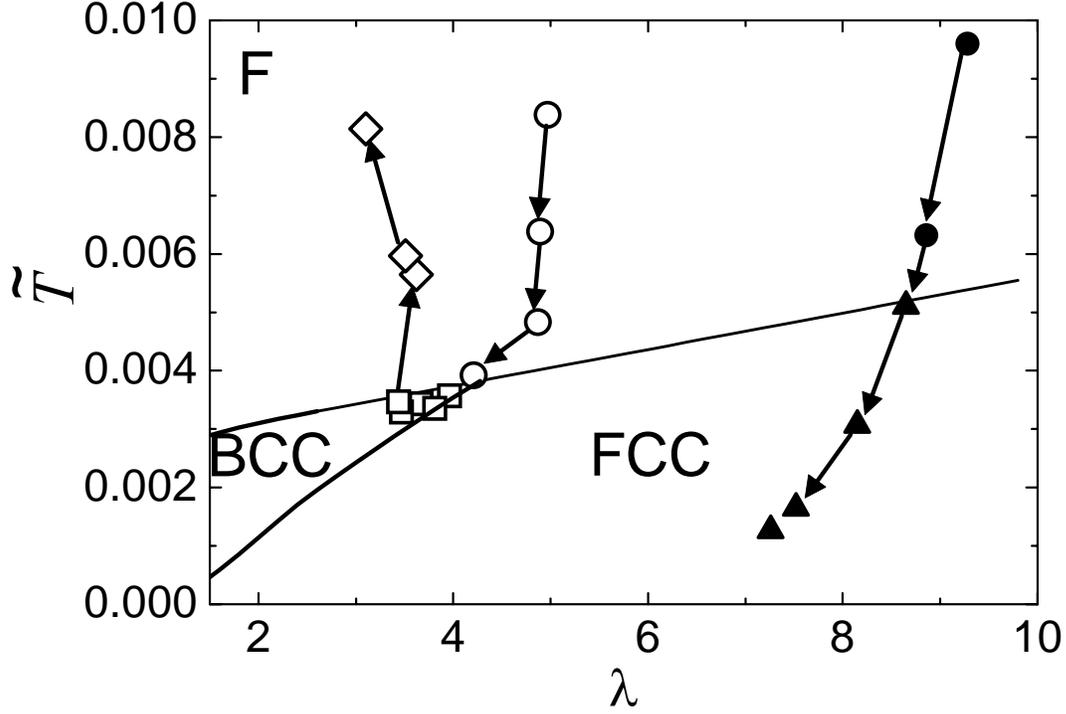

**FIG. 5.** Phase diagram in the $(\lambda, \tilde{T})$ representation (see text). The arrows denote increasing colloid volume fraction. Filled symbols are earlier experimental statepoints,[10] fluid and FCC crystal phases are represented by circles and triangles respectively. The unfilled circles, squares and diamonds correspond to experimental statepoints in the low-density fluid, BCC crystal and re-entrant fluid. The higher density RHCP crystal is not shown in this representation.

| Experiment | CHB component of solvent | Freezing | Re-entrant melting | Solvent conductivity (pS/cm) | Solvent+$\eta$=0.03 colloids conductivity (pS/cm) | Solvent treatment | Ionic strength (M) |
|---|---|---|---|---|---|---|---|
| (1) | CHB 'A' | F-FCC $\eta$=0.1 | none | 2000* | - | Washed+distilled | $4.3 \times 10^{-8}$ |
| (2) | CHB 'B' | F-BCC $\eta$=0.04 | BCC-F, $\eta$=0.11 | 190±30 | 70±5 | Washed+distilled | $4.3 \times 10^{-9}$ |
| (3) | CHB 'C' | F-BCC $\eta$=0.04 | none | 40±5 | 55±5 | Washed | $9.8 \times 10^{-10}$ |
| (4) | CHB 'C' + TBAB salt | F-BCC $\eta$=0.04 | BCC-F $\eta$=0.13 | 190±5 | 115±10 | Washed | |

\* **Measured for pure CHB, not mixture of CHB-decalin.**

**Table I. Phase behaviour using different batches of the CHB solvent component and after addition of TBAB. See text for the ionic strengths in the case of added salt. Note that in the case of CHB 'A', the conductivity was measured for pure CHB, rather than the CHB-decalin mixture.**